\documentclass[12pt]{iopart}

\begin{document}

\title{Phase diagram of a driven interacting three-state lattice gas}
\author{E Lyman and B Schmittmann}
\address{Center for Stochastic Processes in Science and Engineering, and Physics Department, Virginia Tech, Blacksburg, VA 24061-0435, USA}

\date{today}
\vspace{-0.5cm}
\begin{abstract} We present Monte Carlo simulations of a three-state lattice
gas, half-filled with two types of particles which attract one another,
irrespective of their identities. A bias drives the two particle species in
opposite directions, establishing and maintaining a non-equilibrium steady
state. We map out the phase diagram at fixed bias, as a function of
temperature and fraction of the second species. As the temperature is
lowered, a continuous transition occurs, from a disordered homogeneous into
two distinct strip-like ordered phases. Which of the latter is selected
depends on the admixture of the second species. A first order line separates
the two ordered states at lower temperatures, emerging from the continuous
line at a non-equilibrium bicritical point. For intermediate fraction of the
second species, all three phases can be observed.
\end{abstract}
\vspace{-0.7cm}



{\em Introduction. }For systems in thermal equilibrium, the theoretical
framework is firmly laid, resting on the work of Boltzmann and Gibbs over a
hundred years ago. In particular, the study of simple equilibrium models has
a long and illustrious history, as reduction in complexity facilitates the
development of theoretical techniques and intuition. In contrast, for
systems far from equilibrium there exists no general theoretical framework,
and the field remains in an undeveloped state. The strategy of investigating
simple models motivates our Monte Carlo study of a driven diffusive system
far from equilibrium. A modification of the Ising model, our system departs
from well-travelled ground in equilibrium statistical mechanics. Our goal is
to develop some intuition about systems far from equilibrium while extending
earlier work in the field \cite{SZ-rev}.

Almost twenty years ago, Katz, Lebowitz, and Spohn (KLS) \cite{KLS}
introduced a generalization of the Ising lattice gas \cite{Ising}, motivated
by the physics of fast ionic conductors \cite{FIC}: A bias $E$
is applied along a specified lattice axis, driving the particles much like
an electric field would drive positive charges. With conserved density and 
periodic boundary
conditions, the system settles into a non-equilibrium steady state,
characterized by a uniform particle current. Similar to the equilibrium
Ising model, the KLS phase space consists of a high temperature disordered
phase and a low temperature phase-separated phase, characterized by a
particle-rich strip {\em parallel} to the field direction. At half-filling,
the transition remains continuous, but shifts to a higher temperature $%
T_{c}(E)$. Remarkably, the transition belongs to a novel universality
class \cite{FT,MC-KLS,AFSS}, distinct from the Ising class. One of its key
signals is strong anisotropy: wave vectors scale as $k_{\Vert }\sim k_{\bot
}^{1+\Delta }$, with $k_{\Vert }$ in the bias direction and $\Delta =2$ in
dimension $d=2$. Away from $T_{c}(E)$, a conserved order parameter, coupled
with the lifting of the detailed balance constraint, generates power law
decays of correlations at {\em all} $T>T_{c}$ \cite{2-pt}.

A natural generalization \cite{SHZ} of the KLS model introduces a second,
negatively ``charged'' particle species, driven in the opposite direction by
the bias. In the high $T$, high $E$ limit where interparticle interactions
can be neglected, the system has
a line of phase transitions, separating a disordered phase from an
inhomogeneous ordered phase. In the ordered phase, which prevails at high
density, particles of opposite charge block each other's progress,
forming a charge-segregated strip{\em \ transverse} to the field. Both first
and second order transitions can occur \cite{VZS,KSZ}. The blocking
transition persists in systems carrying {\em nonzero} charge, giving rise to
slowly drifting strips \cite{LZ,BZ}. Slow and fast cars, observed in a
co-moving frame, offer a good analogy: a blocking transition (traffic jam)
occurs when vehicles are sufficiently dense \cite{traffic}. We refer to this
noninteracting system as the `two-species model' for short.

It is natural to wonder what will happen if we lift the high field, high
temperature constraint. Now the particles should ``feel'' the Ising
interaction over some range of temperatures, and all three phases (disorder,
parallel and transverse strips) may exist in phase space. Several questions
emerge immediately. First, we can explore the stability of the KLS
universality class by replacing a few positive particles by negative ones.
Eventually, however, a blocking transition will occur when a critical charge
is exceeded. Similarly, the two-species limit can be probed by taking the
strength of interparticle interactions to zero. In this letter, we limit
ourselves to establishing the {\em presence of all three phases}: the
disordered phase, the transverse strip associated with the blocking
transition, and the parallel strip associated with KLS order. Details will
be deferred to a future publication \cite{LS2}. In the next section, we will
introduce the model specifications and our choice of order parameters, to set
the stage for our Monte Carlo results. We conclude with some open questions.

{\em The microscopic model and order parameters.} The configurations of
our model are specified by a set of occupation variables, $\left\{ s\left( 
{\bf r}\right) \right\} $, where ${\bf r}\equiv \left( x,y\right) $ labels a
site on a fully periodic square lattice of dimensions $L_{x}\times L_{y}$,
and each $s\left( {\bf r}\right) $ can take the values $+1$, $-1$, or $0$
for a positive particle, negative particle, or hole. The drive  
$E$ points in the positive $y$-direction. We also introduce the variable 
$n\left( {\bf r}\right) \equiv \left| s\left( {\bf r}\right) \right| $ to
distinguish particles (of either species) from holes. For later reference,
we define the mass density $m=\frac{1}{L_{x}L_{y}}\sum_{{\bf r}}n\left( {\bf %
r}\right) $ and the charge density $q=\frac{1}{L_{x}L_{y}}\sum_{{\bf r}%
}s\left( {\bf r}\right) $. To ensure access to the KLS critical
point, we study systems with $m=0.5$, i.e., half-filled lattices.
The particles are endowed with attractive nearest-neighbor interactions of
strength $J>0$, which are {\em independent }of charge and controlled by the
usual Ising Hamiltonian 
\begin{equation}
H=-4J\sum\limits_{\left\langle {\bf r,r}^{\prime }\right\rangle }n\left( 
{\bf r}\right) n\left( {\bf r}^{\prime }\right)   \label{Ising-H}
\end{equation}
We may set $J=1$ without losing any interesting physics. A given
configuration evolves in time as follows. A nearest-neighbor bond is chosen
at random, and, if occupied by a particle-hole pair, its contents are
exchanged according to the Metropolis \cite{MRRTT} rate $\min \left\{ 1,\exp
\left[ -\left( \Delta H-\delta yEs\left( {\bf r}\right) \right) /T\right]
\right\} $. Here, the second term models the effect of
the drive: if the particle, of charge $s$, is initially located at ${\bf r}$%
, $\delta y$ is the change in its $y$-coordinate due to the jump. Thus,
positive (negative) charges jump preferentially along (against) the field
direction. The parameter $T$ (``temperature'') models the coupling to a
thermal bath. Particle-particle (i.e., charge) exchanges are not allowed.

We note, first, that this dynamics is {\em diffusive}, i.e., it conserves
particle and charge densities. Second, even though the drive mimics an
electrostatic potential, the boundary conditions prohibit the existence of a 
{\em global} Hamiltonian. As a consequence, the system settles into a
generic {\em nonequilibrium} steady state. Third, we briefly review the
different limits of this model: For $q=m=0.5$, we obtain the KLS model,
while $J/T\rightarrow 0$ at finite $E/T$ is the two species case. Of course,
the equilibrium Ising model is recovered for $E=0$, $q=0.5$.
Thus, the natural control parameters for our study are temperature $T$
(measured in units of the Onsager value), the drive $E$ (measured in units
of $J$) and the system charge $q$. 

Due to the conservation laws, the ordered phases are spatially
inhomogeneous. Anticipating strip-like ordered domains, we select an
order parameter sensitive to such structures, i.e., the equal-time
structure factor associated with the particle distribution,
\begin{equation}
\left\langle \Phi (m_{x},m_{y})\right\rangle \equiv \left\langle \left| 
\frac{\pi}{L_{x}L_{y}}\sum_{x,y}n(x,y)e^{2\pi i(m_{x}x/L_{x}+m_{y}y/L_{y})}%
\right| ^{2}\right\rangle   \label{SF}
\end{equation}
Here, $\left\langle \cdot \right\rangle $ denotes a configurational average,
and the integers $m_{x},m_{y}$ index the wave vector. 
Strips transverse and parallel to the drive are
easily identified by considering the smallest nonzero wavevectors in the $x$%
- and $y$-directions, respectively. Specifically, a perfect strip along the $%
y$-direction corresponds to $\left\langle \Phi (1,0)\right\rangle =1$ while
a random configuration gives $\left\langle \Phi \right\rangle =O(\frac{1}{%
L_{x}L_{y}})$. By disregarding the phase, the fluctuations in the strip
position do not interfere with the averaging procedure. While other choices of order parameter are of
course possible, we prefer $\left\langle \Phi (m_{x},m_{y})\right\rangle $
since both its high- and low-temperature limits are exactly known. 
All simulations are run on $40\times 40$ lattices, starting
from random initial configurations except where noted. 
One Monte Carlo step (MCS) is defined as 
$2L_{x}L_{y}$ update attempts. When averaging, the first $2\times 10^{5}$ MCS 
are discarded
to let the system reach the steady state, and measurements are taken every $%
200$ MCS for the next $8\times 10^{5}$ MCS.

{\em Monte Carlo results. }The parameter space for our model is spanned by $T
$, $E$, and $q$. In order to establish the presence of all three phases, $E$
must be chosen judiciously. To date, the driven Ising model has mostly been
studied at infinite drive, where jumps against $E$ are completely
suppressed, in order to maximize its nonequilibrium characteristics. This
choice, however, renders our two-species system non-ergodic: any
configuration in which the minority species forms a blockage, even if it is
just a single row spanning the system in the transverse direction, will
never break up, regardless of whether such a configuration is stable,
metastable, or just a random fluctuation. Thus, a much smaller $E$ must be
selected if we wish to observe transitions from transverse to parallel
strips, with reasonable decay times. Exploratory runs show $E=2$ to be a
good choice. The remaining parameter space is now two-dimensional with axes $%
(T,q)$, and we map out the phase diagram in this plane. We first consider $%
q=0.5$, corresponding to {\em zero} negative charges, at finite $E$. On this
line, we find a single continuous transition, at $T_{c}(2)\simeq 1.15$, to
the KLS ordered state, i.e., a single strip aligned with $E$. A detailed
anisotropic finite-size scaling \cite{AFSS} study, to be published elsewhere \cite{LS2},
indicates that this transition is in the usual KLS class, consistent with
field-theoretic predictions \cite{FT}. Next, we turn to a smaller charge, $%
q=0.425$, which corresponds to exactly $60$ negative particles, i.e., $1.5$
rows, on a half-filled $40\times 40$ lattice. Fig.~1 shows time traces of $%
\Phi \left( 1,0\right) $ and $\Phi \left( 0,1\right) $ at $4$ different
temperatures, all starting from a random initial configuration. 
At high temperatures, the system is disordered, with both modes essentially
zero. As $T$ is lowered to $T\simeq 1.80$, the {\em blocking
transition} occurs first, evidenced by $\Phi \left( 0,1\right) $ emerging
from the noise. At $T=1.00$,  
the system settles quickly into a well-developed transverse strip 
with a strong nonzero signal
in $\Phi \left( 0,1\right) $. Lowering $T$ further to $T=0.90$, we
observe signatures of metastability: the corresponding time trace 
initially develops a large $\Phi \left( 1,0 \right) $ which decays after
about $2.5\times 10^{5}$ MCS and reorganizes itself into a transverse strip, 
signalled by $\Phi \left( 0,1 \right) $. Finally, a
well-developed parallel strip is observed at $T=0.80$. 
Thus, we establish a sequence of {\em two
transitions} at $q=0.425$, with the blocking transition occuring {\em first}
as the temperature is lowered.

\begin{figure}[tbp]
\input{epsf}
\begin{center}
\vspace{-2.cm}
\begin{minipage}{0.6\textwidth}
  \epsfxsize = \textwidth \epsfysize = 0.8\textwidth \hfill
  \epsfbox{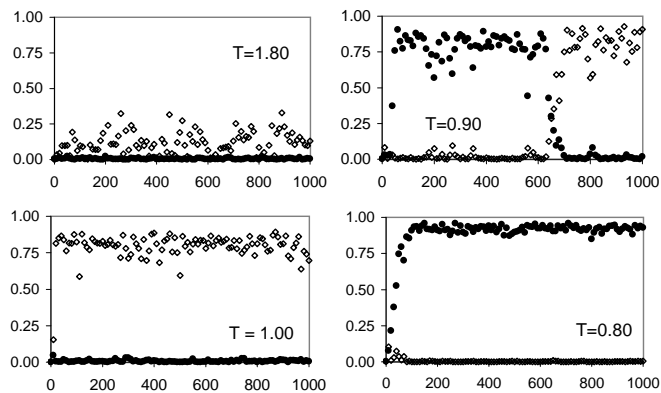}
    \vspace{-1.cm}
\end{minipage}
\end{center}
\caption{Time traces for $\Phi \left( 0,1\right) $ ($ \diamond $) and 
$\Phi \left( 1,0\right) $ ($\bullet $) vs time (in units of 400 MCS), 
at four different temperatures for $q=0.425$.} 
\vspace{-0.5cm}
\end{figure}

The results of our simulation study are summarized in Fig.~2 which shows the
phase diagram in the $q$-$T$ plane, for a $40\times 40$ system. As $q$ {\em %
decreases} from $0.5$ to $0$, the number of negatively charged particles 
{\em increases} from $0$ to $400$. To locate and distinguish continuous and
first order transitions, we monitor time traces of the order parameters and
extract their fluctuations, i.e., $\left\langle \Phi ^{2}\left( 1,0\right)
\right\rangle -\left\langle \Phi \left( 1,0\right) \right\rangle ^{2}$ and $%
\left\langle \Phi ^{2}\left( 0,1\right) \right\rangle -\left\langle \Phi
\left( 0,1\right) \right\rangle ^{2}$. Crossing a continuous transition, the
appropriate order parameter rises smoothly from the noise, accompanied by a
peak in its fluctuations. For example, at $q=0.425$, the fluctuations of
the $\left( 0,1\right) $ mode peak at $T=1.80\pm 0.05$, whence we use this
value to (approximately) locate the transition. Proceeding in this manner,
we find two lines of continuous transitions, separating the disordered phase
from two different ordered phases: For $0.50\geq q\geq $ $0.46$, the
disordered (D) phase becomes unstable with respect to a {\em parallel} 
 strip (PS)
 as in the KLS model, while for $q\leq $ $0.45$ the system orders into
the {\em transverse} strip (TS) associated with the blocking transition.
However, the parallel strip reemerges, as the true low-temperature
configuration: A line of first order transitions begins at $q\simeq 0.46$, $%
T\simeq 1.1$, extending to smaller $q$'s and $T$'s. This line separates two 
{\em ordered} phases: transverse strips which persist at higher
temperatures, and parallel strips at lower $T$'s. Near this line, time
traces of the order parameters show metastability and hysteresis. To
separate stable from metastable configurations (to the accuracy of our
simulations), we analyzed long runs up to $2.4\times 10^{6}$ MCS,
starting from different initial configurations. For $%
q=0.425$, a sharp transition is easily located at $T=0.84\pm 0.01$. This
becomes more difficult as $q$ increases since the continuous and first order
lines approach one another and the first order character of the lower
transition weakens. For smaller $q$, the first order transition shifts to
such low temperatures that metastable states are effectively frozen on the
time scales of our simulations.

\begin{figure}[tbp]
\input{epsf}
\begin{center}
\vspace{-1.5cm}
\begin{minipage}{0.4\textwidth}
  \epsfxsize = \textwidth \epsfysize = 0.8\textwidth \hfill
  \epsfbox{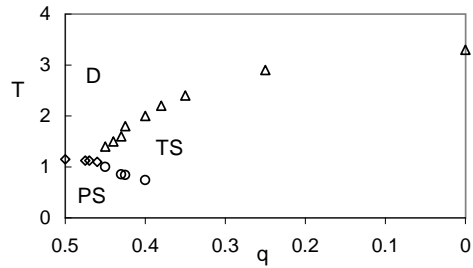}
    \vspace{-1.2cm}
\end{minipage}
\end{center}
\caption{Phase diagram for a $ 40 \times 40 $ system in the $q,T$ plane, 
for $E=2 $.
The D-PS ($ \diamond $) and D-TS ($ \triangle $) transition lines
are second order. The PS and TS phases
are separated by first order transitions ($ \circ $). The junction of the three
lines marks the bicritical point.}
\vspace{-0.5cm}
\end{figure}

Provided our qualitative picture is confirmed by further tests, the junction
of the first and second order lines, at $q\simeq 0.46$, $T\simeq 1.1,$ is a 
{\em non-equilibrium bicritical} point. In fact, the system shows markedly
different behavior at $q=0.45$ and $q=0.46$. At $q=0.46$ the order parameter
signalling transverse strips, $\left\langle \Phi \left( 0,1\right)
\right\rangle $, reaches a maximum value of just $0.09$ at $T=1.2$, in stark
contrast to a maximum value of $0.33$ at $T=1.1$ if $q=0.45$. 
Moreover, 
the $q=0.45$ system clearly shows a lower transition 
with signs of metastability, 
while none
is observed at $q=0.46$. While it is quite remarkable that changing the sign
of just {\em eight} particles makes such a difference, we emphasize that $%
q=0.45$ corresponds to precisely {\em one full row} of negative particles in
a $40\times 40$ system while $q=0.46$ results only in a partially filled row
($32$ negative charges).

{\em Conclusions. }We have simulated a lattice gas, consisting of two
oppositely charged particle species and holes subject to 
an ``electric'' field. The particles attract one
another, independent of charge. This model interpolates between two
well-studied limits: the KLS model \cite{KLS} which has just a single
species, and the high-field, high-temperature version of this model \cite
{SHZ} where the interactions are irrelevant. Both limits exhibit
order-disorder transitions, characterized however, by different ordered
phases: a density-segregated strip {\em parallel }to the drive in the KLS
limit, and a density- and charge-segregated strip {\em transverse} to $E$ for the
two-species limit. Here, we have mapped out the phase diagram for an
intermediate value of the drive, where both ordered phases are observed in
different regions of parameter space. Lines of first and second order
transitions, joined at a bicritical point, demarkate their stability domains.

We conclude with some remarks on work in progress \cite{LS2} and open
questions. Clearly, we have explored only a limited portion of the huge
parameter space.  Moreover, a systematic finite-size scaling analysis is
needed to obtain a better estimate of the transition lines and to extract
the critical properties of the continuous transitions. Preliminary studies
show that the mean features of our phase diagram are independent of system
size. Analytic work, ranging from mean-field to full-fledged renormalized
field theory, should provide further insights. A particularly intriguing
question is how $q$ and $E$ should scale with the lattice dimensions. If
both are held fixed when performing the standard finite size analysis for
the KLS model, the blocking transition will eventually supersede the KLS
transition: a fixed charge density corresponding to a single row in a square
system becomes several rows thick in a ``long skinny'' system. Clearly, much
remains to be explored before this rich system is understood in detail.

{\em Acknowledgements.} We thank R.K.P.~Zia, R.J.~Astalos and U.C.~T\"{a}uber for helpful discussions. Partial support from the National Science Foundation through DMR-0088451 is gratefully acknowledged.

{ \bf References}

\end{document}